\documentclass[prl,superscriptaddress,amsmath,amssymb,showpacs,noshowkeys,a4paper,twocolumn]{revtex4}

\usepackage{graphicx}% Include figure files
\usepackage{dcolumn}% Align table columns on decimal point
\usepackage{bm}% bold math
\usepackage[usenames]{color}
\usepackage{epstopdf}
\usepackage{wasysym}
\definecolor{mygray}{gray}{0.5}

\newcommand{\affpof}{\affiliation{Physics of Fluids Group, MESA+ Institute for Nanotechnology, J. M. Burgers Centre for Fluid Dynamics, University of Twente, P. O. Box 217, 7500 AE Enschede, The Netherlands}}

\begin{document}

\title{Symmetric and Asymmetric Coalescence of Drops on a Substrate}
\author{J.F.~Hern\'andez-S\'anchez}
\affpof
\author{L.A.~Lubbers}
\affpof
\author{A.~Eddi}
\affpof
\author{J.H.~Snoeijer}
\affpof

\begin{abstract}
The coalescence of viscous drops on a substrate is studied experimentally and theoretically. We consider cases where the drops can have different contact angles, leading to a very asymmetric coalescence process. Side view experiments reveal that the ``bridge" connecting the drops evolves with self-similar dynamics, providing a new perspective on the coalescence of sessile drops. We show that the universal shape of the bridge is accurately described by similarity solutions of the one-dimensional lubrication equation. Our theory predicts a bridge that grows linearly in time and stresses the strong dependence on the contact angles. Without any adjustable parameters, we find quantitative agreement with all experimental observations.

\end{abstract}
\pacs{47 55.D- Drops}

\maketitle

The coalescence or breakup of liquid drops is a fundamental process relevant for the formation of raindrops or sprays, inkjet printing, or stability of foams and emulsions \cite{EggersRMP1997,KapurPRE2007,AndrieuBeysensJFM2002}. The initial stages of coalescence of two spherical drops has been characterized in great detail \cite{BrennerPRL1994,EggersJFM1999, HopperJFM1990,ThoroddsenJFM2005, AartsPRL2005,DucheminJFM2003,PaulsenPRL2011}. After contact, a small liquid bridge connects the two drops and the bridge grows rapidly with time. Depending on the viscosity of the liquid, the radius of the bridge grows as $r \sim t$ (high viscosity) \cite{EggersJFM1999, HopperJFM1990,ThoroddsenJFM2005, AartsPRL2005}, or $r \sim t^{1/2}$ (low viscosity, inertia dominated) \cite{DucheminJFM2003, ThoroddsenJFM2005, AartsPRL2005}, with a crossover depending on fluid properties and drop size \cite{PaulsenPRL2011}.

In many cases, however, the coalescing drops are not freely suspended but are in contact with a substrate. Much less is known about the coalescence of such sessile drops. When looking from a top view (perpendicular to the substrate), the coalescence of drops on a substrate looks very similar to the case for spherical drops \cite{AndrieuBeysensJFM2002}; yet the bridge dynamics is fundamentally different. Measurements for very viscous drops give a growth $r \sim t^{1/2}$ \cite{RistenpartPRL2006, NarheEPL2008}, and even smaller exponents have been suggested \cite{YarinLANGMUIR2012}. The challenge lies in the complications introduced by the presence of the substrate. First, the geometry of the drop is no longer a sphere with an axisymmetric bridge, but a spherical cap with a contact angle $\theta$. As a consequence, a top view of the coalescence process is very different from a side view. Second, the wall slows down the liquid transport towards the bridge \cite{RistenpartPRL2006} and gives rise to the motion of a contact line \cite{BonnRMP2009}. At present, it is not clear whether or not this contact line motion affects the initial stages of coalescence, and different predictions for the $\theta$ dependence have been reported \cite{RistenpartPRL2006, NarheEPL2008,YarinLANGMUIR2012}.

\begin{figure}[t]
\includegraphics[width=1\columnwidth]{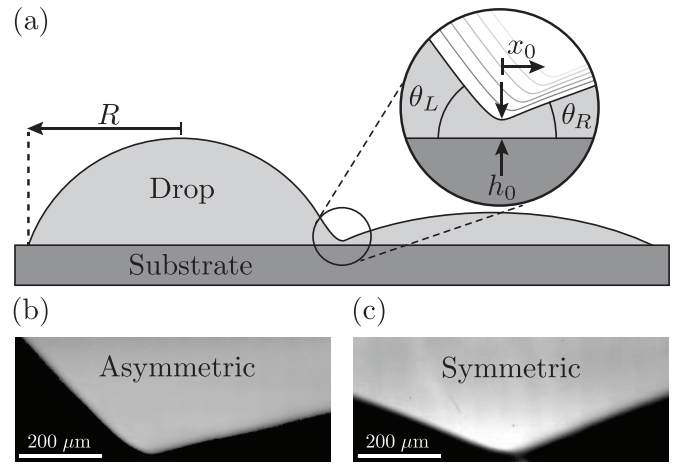}
\caption{\label{fig:ske} (a) Schematic of two coalescing viscous drops on a substrate, viewed from the side. The minimum height $h_0(t)$ characterizes the bridge size. The left-right contact angles $\theta_L$ and $\theta_R$ can be different at the moment of contact. The horizontal displacement $x_0$ results from the asymmetry in the contact angles. (b,c) Typical frames of the experiments are shown for asymmetric contact angles (b) and symmetric contact angles (c).}
\end{figure}

In this Letter we resolve the coalescence of viscous drops on a substrate by performing side view experiments, imaging parallel to the substrate (Fig.~\ref{fig:ske}). Our central finding is that the initial stages evolve by a self-similar shape of the bridge, with a linear growth of the bridge height $h_0 \sim t$. The influence of the contact angle is studied in detail by considering drops with identical or different contact angles, resulting into symmetric or asymmetric coalescence [Fig.~\ref{fig:ske}(bc)]. 
Theoretically, we show that all experiments can be described quantitatively by a similarity solution of the lubrication equation. Remarkably, this one-dimensional approach quantitatively predicts the shape and time evolution of the bridge without adjustable parameters. Our results reveal that the rate of vertical growth scales with the contact angle as $\sim \theta^4$, the horizontal speed $\sim \theta^3$, and provide a new perspective on previous top view measurements.

\begin{figure}[t]
\includegraphics[width=1\columnwidth]{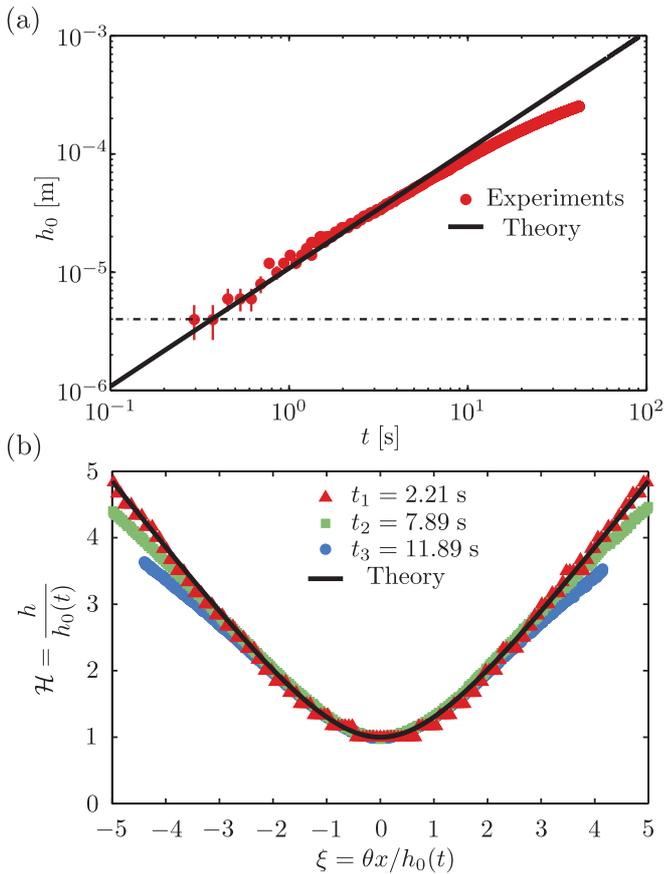}\\
\caption{\label{fig2} (color online) Symmetric coalescence. (a) Height of the bridge $h_0$ as a function of time after contact $t$, for drops with $\theta_L=\theta_R=22^{\circ}$ ($\eta=12.2\mathrm{\;Pa \cdot s}$). Experiments are shown in red ($\CIRCLE$), the solid line is the prediction by Eqs.~(\ref{eq:sim_sol},\ref{eq:speed}). The dashed line represents the lower limit for spatial resolution.
(b) Rescaled experimental profiles at different times, ${\cal H}=h(x,t)/h_0(t)$ versus $\xi = x\theta/h_0(t)$. The collapse reveals self-similar dynamics at the early stage of coalescence, in agreement with the similarity solution obtained from our analysis (solid line). 
}
\end{figure}

\paragraph{Experimental setup.---} The side view images of coalescing drops in Fig.~\ref{fig:ske}(bc) are obtained by a digital video camera (Photron APX-RS) equipped with a microscopic lens (Navitar 12x zoom lens), resulting in a resolution of $2\;\mu \mbox{m/pixel}$. The camera recorded $12.5$ frames per second. The substrate consists of a horizontal microscope glass slide (Menzel pre cleaned microscope slide, average roughness $\approx10\; \mbox{nm}$). The glass slide was further cleaned using ethanol and acetone, then submerged in an ultrasonic bath and dried with filtered nitrogen gas. The coalescing drops were made from silicon oils (Basildon Chemical Company Limited), with viscosity $\eta = 0.974\mathrm{\;Pa \cdot s\;}$ or $12.2\mathrm{\;Pa \cdot s}$, which both have a  surface tension $\gamma=21\cdot10^{-3}\mathrm{\;N \cdot m}^{-1}$ and density $\rho = 975 \mathrm{\;kg \cdot m^{-3}}$. The silicon oils perfectly wet the cleaned glass slide ($\theta_{eq} \approx 0$). 

The coalescence of two drops is controlled as follows. A first drop is deposited from the syringe on the substrate. Although the silicon oil perfectly wets the glass, the spreading of these high viscosity drops is very slow, with the liquid contact angle decreasing slowly in time. Subsequently, the glass plate is displaced  by a manual translation stage and a second drop is placed next to the first one. By controlling the expelled volume of silicon oil and the time between the deposition of drops we achieve a range of contact angles $\theta_L$ and $\theta_R$ between $10^{\circ} \mbox{ to } 67^{\circ}$ at the time of coalescence. We consider both symmetric coalescence [$\theta_L=\theta_R$, Fig.~\ref{fig:ske}(c)] and asymmetric coalescence [$\theta_L\neq \theta_R$, Fig.~\ref{fig:ske}(b)]. The spreading determines the initial conditions, but in all cases the spreading speed is much smaller than the bridge growing speed. Contact time is determined when there is a visual change, which happens before the bridge is thick enough to provide a reliable measurement. The dashed line in Fig.~\ref{fig2}(a) shows this spatial resolution limit.

\paragraph{Self-similar dynamics.---} The dynamics of coalescence is characterized by the growth of the bridge connecting the two drops. Figure~\ref{fig2}(a) presents the minimum height of the bridge, $h_0$, as a function of time for a symmetric coalescence experiment ($\theta_L=\theta_R= 22^{\circ}$). At early times, we observe a linear increase of the bridge height, i.e. $h_{0}\sim t$, while at later times the coalescence slows down. In these final stages the height of the bridge becomes comparable to the total drop size, which is typically $\sim 1~$mm for all experiments. The very early stage, however, exhibits self-similar dynamics that is governed by a single length scale. This is revealed in Fig.~\ref{fig2}(b) where the meniscus profiles, $h(x,t)$, and the horizontal coordinate, $x$, are rescaled by $h_0(t)$. The scaled profiles at different times collapse onto a universal curve: the early stages of coalescence are characterized by a self-similar meniscus profile. The size of the bridge is simply $h_0$, both in horizontal and vertical direction. The solid line is the theoretical similarity profile that will be derived below.

Our experiments suggest that coalescence of drops on a substrate is governed by a similarity solution of the flow. To simplify the three-dimensional geometry of the coalescence, we assume that the flow is predominantly oriented from the centers of the drops towards the coalescing bridge, as suggested by Ristenpart {\it et al.}~\cite{RistenpartPRL2006}. We therefore attempt a similarity solution based on the one-dimensional lubrication theory for viscous flows~\cite{OronRMP1997}: 
\begin{equation}
	\frac{\partial h}{\partial t} + \frac{\gamma}{3\eta}\frac{\partial}{\partial x} \left( h^3\frac{{\partial}^3 h}{\partial x^3} \right)=0.
	\label{eq:lub_appr}
\end{equation}
Here, $h(x,t)$ is the meniscus profile viewed from the side, $\eta$ is the liquid viscosity and $\gamma$ denotes the surface tension. This lubrication equation is valid for small contact angles and represents mass conservation: the second term is the surface tension-driven flux of liquid towards the bridge, causing a growth of the bridge ($\partial h/\partial t > 0$). 

Consistent with our experiments, Eq.~(\ref{eq:lub_appr}) has a similarity solution that imposes a linear time-dependence,

\begin{equation}
	h\left(x,t \right)= v t \, {\cal H} \left( \xi \right), \quad \mbox{with} \quad \xi = \frac{\theta x}{vt},
	\label{eq:sim_sol}
\end{equation}
where ${\cal H}(\xi)$ is the similarity profile of the meniscus bridge. Here we incorporated the contact angle $\theta$ in the scaling of $x$, such that the condition $\partial h/\partial x = \theta$ translates to ${\cal H}' = 1$. The correct scaling of the coalescence velocity with $\theta$ then turns out to be

\begin{equation}\label{eq:speed}
v = V \frac{\gamma \theta^4}{3 \eta},
\end{equation}
where $V$ is a numerical constant that still needs to be determined. In combination with (\ref{eq:lub_appr}) and (\ref{eq:sim_sol}), this provides an ODE for the similarity profile ${\cal H}(\xi)$:

\begin{equation}
	{\cal H} - \xi {\cal H}' + \frac{1}{V}\left( {\cal H}^3{\cal H'''} \right)'=0.
	\label{eq:ode}
\end{equation}
In order to solve Eq.~(\ref{eq:ode}), which is a fourth order ODE with one unknown parameter $V$, five boundary conditions are required. At the center of the symmetric bridge

\begin{equation}\label{eq:BC1}
{\cal H}(0)=1, \quad {\cal H}'(0)={\cal H}'''(0)=0,
\end{equation}
while far away the profile has to match a linear slope of contact angle $\theta$. For the similarity variables this becomes

\begin{equation}\label{eq:BC2}
{\cal H}''(\infty)=0, \quad {\cal H}'(\infty)=1.
\end{equation}
The boundary value problem (\ref{eq:ode}-\ref{eq:BC2}) uniquely determines the similarity solution for symmetric drop coalescence. It was solved numerically using a shooting algorithm, from which we obtained both the dimensionless velocity, $V = 0.818809$, and the similarity profile ${\cal H}(\xi)$. As the influence of the contact angle was scaled out, the solution describes the coalescence for all contact angles, within the lubrication assumption of small $\theta$.

The similarity solution indeed provides an accurate description of the coalescence experiments. The solid line in Fig.~\ref{fig2}(a) is the prediction (\ref{eq:speed}) without adjustable parameters. The solid line in Fig.~\ref{fig2}(b) is the similarity profile ${\cal H}(\xi)$ obtained from our analysis. The agreement between experiment and theory shows that the initial stages of coalescence are accurately described by a one-dimensional lubrication model. As expected, the similarity solution breaks down at later times when the size of the meniscus bridge becomes comparable to the size of the drops.

\begin{figure}[t]
\includegraphics[width=1\columnwidth]{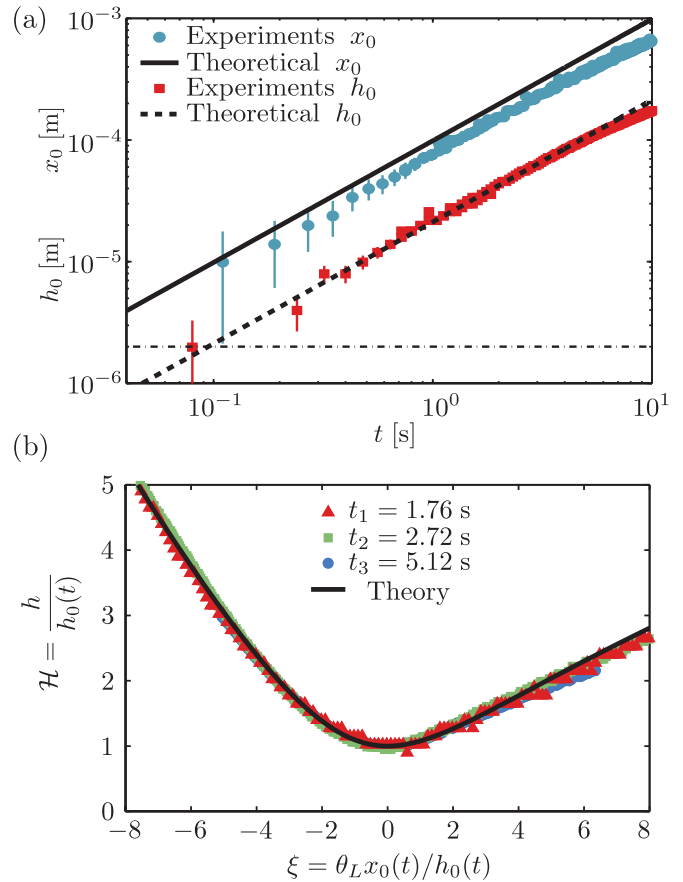}
\caption{\label{fig3} (color online) Asymmetric coalescence. (a) Horizontal  and vertical position of the meniscus bridge, $x_0(t)$ and $h_0(t)$, for asymmetric drops ($\theta_{L}=46^{\circ}$, $\theta_{R}=13^{\circ}$, viscosity $\eta=12.2\mathrm{\;Pa \cdot s}$). Blue ($\CIRCLE$) and red ($\blacksquare$) markers are experimental data for $x_0$ and $h_0$ respectively. The solid and dashed lines are the predictions from the similarity solutions. 
(b) Rescaled experimental profiles at different times, ${\cal H}=h(x,t)/h_0(t)$ versus $\xi = x_0\theta_L/h_0(t)$. The collapse reveals self-similar dynamics at the early stage of coalescence. The solid line is the similarity solution predicted by our analysis.
}
\end{figure}

\begin{figure*}[t]
\includegraphics[width=1\linewidth]{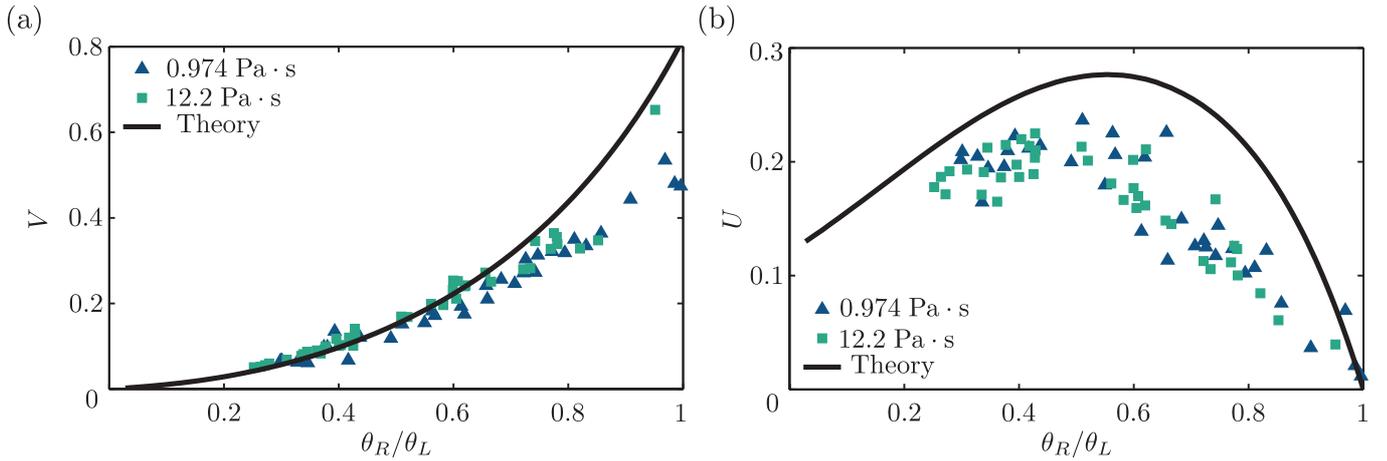}
\caption{\label{fig4} (color online) Contact angle dependence of coalescence velocity. (a) Dimensionless vertical speed, $V=3v\eta/(\gamma \theta_L^4)$, as a function of $\theta_R/\theta_L$. (b) Dimensionless horizontal speed, $U=3u\eta/(\gamma \theta_L^3)$, as a function of $\theta_R/\theta_L$. The horizontal speed vanishes for the symmetric case $\theta_R/\theta_L=1$, and displays a maximum around $\theta_L/\theta_R \sim 0.5$. Symbols correspond to 75 experiments, solid lines are predictions obtained from the similarity solutions.}
\end{figure*}

\paragraph{Asymmetric coalescence.---} We further extend the theory to asymmetric coalescence, for which the contact angles $\theta_L\neq \theta_R$ (Fig~\ref{fig:ske}). Without loss of generality, we assume that $\theta_{L} > \theta_{R}$, and scale the coordinates using $\theta_{L}$. Interestingly, the lack of symmetry induces a \emph{horizontal} displacement of the meniscus bridge during the coalescence process: the minimum of the bridge, $x_0$, is pulled towards the lower contact angle ($\theta_R$). This effect can be captured using a similarity variable that is co-moving with the bridge, of the form

\begin{equation}
	\xi=\frac{\theta_{L}(x-ut)}{vt},  \quad \mbox{with} \quad 
	u= U \frac{\gamma {\theta}_{L}^{3}}{3 \eta}.
	\label{eq:asim_sol}
\end{equation}
The horizontal velocity of coalescence $u$ scales with $\theta_L^3$, where $U$ is a numerical constant. The vertical velocity still follows (\ref{eq:speed}) with $\theta=\theta_L$. Inserting (\ref{eq:asim_sol}) in (\ref{eq:lub_appr}) yields

\begin{equation}
	{\cal H} - \left(\xi + \frac{U}{V}\right) {\cal H}' + \frac{1}{V}\left( {\cal H}^3{\cal H'''} \right)'=0.
	\label{eq:asym_ode}
\end{equation}
This fourth order ODE for ${\cal H}(\xi)$ now contains two unknown parameters, $U$ and $V$, hence the solution requires six boundary conditions. The minimum of the bridge is still defined by ${\cal H}(0)=1$, ${\cal H}'(0)=1$, but the symmetry condition on ${\cal H}'''$ no longer applies. Instead, one has to impose ${\cal H}''(-\infty)={\cal H}''(\infty)=0$, while the asymmetric contact angles give

\begin{equation}
 \quad {\cal H}'(- \infty)=- 1, \quad {\cal H}'(\infty) = \theta_R/\theta_L. 
\end{equation}
With these conditions the boundary value problem has a unique solution for each ratio of contact angles $\theta_R/\theta_L$. This means that $U$, $V$ and ${\cal H}(\xi)$ can again be determined numerically for arbitrary $\theta_R/\theta_L$.

Figure~\ref{fig3} compares theory and experiment for an asymmetric coalescence ($\theta_R/\theta_L=0.25$). The horizontal position of the bridge $x_0$ (blue circles) and the vertical position of the bridge $h_0$ (red squares) are shown in Fig.~\ref{fig3}(a). These again evolve linearly in time with a well-defined velocity. The solid and dashed lines are the predictions (\ref{eq:speed},\ref{eq:asim_sol}), with prefactors $U$ and $V$ determined from the similarity solution. Figure~\ref{fig3}(b) confirms that the asymmetric experimental profiles indeed display self-similarity (symbols), in excellent agreement with theory (solid line).

We finally consider the influence of the contact angle on the coalescence speed. Our theory suggests a universal behavior when making the horizontal and vertical velocities dimensionless, according to $U=3u\eta/(\gamma \theta_L^3)$ and $V=3v\eta/(\gamma \theta_L^4)$. The results of 75 experiments are summarized in Fig.~\ref{fig4} and compared to the theoretical prediction. Indeed, we observe a good collapse of the data. An interesting feature is that the theory predicts an optimal horizontal speed around $\theta_R/\theta_L\approx 0.5$, which is verified experimentally (although the experimental velocities are slightly smaller than expected). This maximum horizontal velocity can be explained as follows. The asymmetry induces a bias in the pulling force of surface tension, which is more efficient for the smaller contact angle $\theta_R$. However, the ``lubrication effect'' inhibits liquid transport when $\theta_R\rightarrow 0$, as the viscous friction in the liquid increases for smaller angles. The combination of these two effects gives rise to an optimum ratio $\theta_R/\theta_L$. 

\paragraph{Discussion.---} 

Our results imply that the initial coalescence of drops on a substrate, which is manifests three-dimensionally, is described quantitatively by a one-dimensional model. This can be explained from the cross-section of the bridge perpendicular to our viewpoint, $r$, which is much larger than $h_0$. Elementary geometry suggests $r \sim (R h_0/\theta)^{1/2}$ \cite{NarheEPL2008}, $R$ being the footprint radius of the drop on the substrate, such that indeed $r \gg h_0$ at early times. Combining this geometric relation with (\ref{eq:speed}) gives a growth law for the top view size of the bridge, $r \sim \theta^{3/2} (R\gamma t/\eta)^{1/2}$. The dependence $\sim \theta^{3/2}$ differs from previous predictions~\cite{NarheEPL2008}, but agrees with the data collapse proposed by Ristenpart \emph{et al.}~\cite{RistenpartPRL2006}. It would be interesting to see whether the one-dimensional approach also applies for coalescence of low-viscosity drops, which are dominated by inertia rather than viscosity \cite{BillinghamKingJFM2005,KapurPRE2007}. 

Our findings also highlight the key importance of asymmetry on the coalescence dynamics. While here the asymmetry is due to the contact angles, a similar effect was found for merging drops with unequal surface tensions (such as water and alcohol), for which the coalescence is strongly delayed by Marangoni forces ~\cite{Karpitschka2010}. This will have a strong bearing on applications as inkjet printing, for which such asymmetries are encountered naturally due to spreading and evaporation of ink drops.

\acknowledgements{We thank K. Winkels and Sander Huisman for discussions. This work is sponsored by Lam Research, STW and NWO by VIDI grant N$^\circ$11304.}

%\bibliography{bibliography}

\end{document}